\documentclass{aa}
\usepackage{graphicx}
\usepackage[caption=false]{subfig}
\usepackage{float}
\usepackage{txfonts}
\usepackage{braket}
\usepackage{appendix}
\usepackage{amsmath}
\usepackage{bm}
\usepackage[colorlinks,allcolors=blue]{hyperref}
\usepackage[a]{esvect}
\makeatletter
\renewcommand*\aa@pageof{, page \thepage{} of \pageref*{LastPage}}

\makeatother
\usepackage{amstext}
\usepackage[normalem]{ulem}

\begin{document}

\title{GJ\,2126 b: A highly eccentric Jovian exoplanet}

\author{A. Schorr\inst{\ref{inst1}} \and A. Binnenfeld\inst{\ref{inst2}} 
\and S. Zucker\inst{\ref{inst2},\ref{inst1}}}

\institute{
School of Physics and Astronomy, Raymond and Beverly Sackler Faculty of Exact Sciences, 
Tel Aviv University, Tel Aviv, 6997801, Israel
\label{inst1}
\and
Porter School of the Environment and Earth Sciences, Raymond and Beverly Sackler Faculty of Exact Sciences, 
Tel Aviv University, Tel Aviv, 6997801, Israel \label{inst2}}

\date{Accepted XXX. Received YYY}

\abstract{
We report the discovery of \object{GJ\,2126\,b}, a highly eccentric ($e = 0.85$) Jupiter-like planet orbiting its host star every $272.7$ days. The planet was detected and characterized using $112$ radial velocity (RV) measurements from HARPS (High Accuracy Radial velocity Planet Searcher), provided by HARPS-RVBank. This planet orbits a low-mass star and ranks among the most eccentric exoplanets discovered, placing it in a unique region of the parameter space of the known exoplanet population. This makes it a valuable addition to the exoplanet demographics, helping to refine our understanding of planetary formation and evolution theories.
}


\titlerunning{GJ\,2126 b: A highly eccentric Jovian exoplanet}
\authorrunning{A. Schorr et al.}

\maketitle
\section{Introduction}
\label{sec:intro}

Given the mean eccentricity of $e = 0.06$ for planets in the Solar System, the discovery of exoplanets with highly eccentric orbits has emerged as one of the most surprising findings in exoplanet research \citep[e.g.,][]{2009MNRAS.392..641O}. Highly eccentric planets exhibit significant variations in their distance from their host stars throughout their orbits, suggesting dynamic histories influenced by diverse gravitational interactions \citep{2014IAUS..299.....B}. This opens up the possibility for various proposed formation and evolution processes \citep[][]{2015Winn}, such as resonance interaction in the protoplanetary disk \citep{2007Cresswell}, planet-planet scattering \citep{2014prpl.conf..787D, 2019A&A...629L...7C}, and stellar flybys \citep{2011MNRAS.411..859M}.

Recent studies indicate that eccentric Jupiters, defined as giant exoplanets with  $e > 0.25$, are abundant, as shown by \citet{2018ApJ...856...37B}. This conclusion is based on the population characterized following the discovery of the first eccentric Jupiter \citep[\object{16\,Cyg\,Bb};][]{1997ApJ...483..457C}. However, this does not extend to highly eccentric Jupiters, specifically those with $e > 0.6$, for which both current models and observations indicate only a limited number \citep{2007AJ....134.1276W, 2020A&A...643A..66B}.

This fact is linked to the unique and dramatic evolutionary path required for such systems to develop \citep[e.g.,][]{2018AJ....156..192W}, believed to occur due to planet–planet gravitational interactions following the gas disc dispersal \citep{2008ApJ...686..603J, 2009ApJ...699L..88R}. Most scenarios that could give rise to such systems would likely lead to moderately eccentric orbits or, conversely, result in the complete ejection of the planet \citep[e.g.,][]{2013MNRAS.431.3494L}.

In this work, we report the detection of an exoplanet occupying this unique parameter space: a highly eccentric Jupiter-like exoplanet, orbiting the low-mass star \object{GJ\,2126}. The paper is structured as follows: We first discuss the spectroscopic observations of \object{GJ\,2126} in Sect.~\ref{sec:Observations}. We then describe their analysis in Sect.~\ref{sec:analysis}, where we also present the characteristics of the newly detected planet. Finally, we summarize and discuss our findings in Sect.~\ref{sec:discussion}.

\section{Observations}
\label{sec:Observations}

\subsection{HARPS}
 HARPS \citep[High Accuracy Radial velocity Planet Searcher;][]{2003Msngr.114...20M} is a high-resolution visible-light echelle spectrograph installed at the European Southern Observatory (ESO) $3.6$-m telescope in La Silla, Chile. With its proven long-term stability and simultaneous calibration it has been demonstrated to reach radial-velocity (RV) accuracy of $\sim 1 ~$\text{m s$^{-1}$} \citep[see][]{2014Natur.513..358P}.

Despite its high precision, HARPS data are still known to contain minor residual systematics. For instance, \citet{2015ApJ...808..171D} identified a one-year variability linked to block stitches in the CCD detector, and a fiber upgrade in May 2015 \citep{2015Msngr.162....9L} introduced a spectral-type-dependent shift in the RV zero point, of approximately $ 10 ~\text{m/s}$.

Recently, \citet{2020A&A...636A..74T} obtained improved and more precise RVs using SERVAL \citep[SpEctrum Radial Velocity AnaLyser;][] {2018A&A...609A..12Z}, corrected for HARPS Nightly Zero-Point variations (NZPs; \citealt{Tal-or2019a}), and made the corrected data publicly available as the HARPS-RVBank dataset. \citet{2024Perdelwitz} later introduced an advanced version, offering enhanced precision and extended temporal coverage. We used measurements from this public dataset for the analysis presented in the following sections.

\subsection{Measurements}
\label{sec:Measurements}

\object{GJ\,2126} is a high proper motion star, and its stellar properties are summarized in Table~\ref{tab:table 1}. \object{GJ\,2126} was observed by HARPS over a span of $15$ years (2004–2019, $\sim 5600$ days), with a total of $112$ measurements. The measurements, labeled as DRVmlcnzp in \citet{2024Perdelwitz}, have an average S/N of $45.0$ and a typical RV error of $1.9$  m s$^{-1}$.

Due to the fiber upgrade in May 2015 \citep{2015Msngr.162....9L}, we chose to treat the pre- and post-upgrade RVs as separate data sets from two distinct instruments, each potentially having different zero points (hereafter referred to as "pre-" and "post-"). Of the $112$ measurements of \object{GJ\,2126}, $95$ were obtained before the fiber upgrade, while $17$ were taken afterward.

In 2020, the observatory was temporarily shut down due to the COVID-19 pandemic\footnote{Nov 9, 2020 entry; \url{https://www.eso.org/sci/facilities/lasilla/instruments/harps/news.html}}, which introduced an additional offset in the data due to the instrument's warm-up. As only two further measurements of \object{GJ\,2126} were taken after this event, we decided to exclude them from our analysis. No additional data points were omitted.

Although its effective temperature of $4159 \pm 129$ K suggests it is likely a K-type star, \object{GJ\,2126} is commonly referred to in the literature as an M-dwarf \citep[e.g.,][]{2019ApJ...878..134K, 2023Mignon}, a designation that traces back to \citet{1975AJ.....80..972S}. While this classification may require reconsideration, it does not affect the results of this paper.

Red dwarfs, of spectral types M and late K, typically exhibit significant stellar activity \citep[see][]{2023Mignon} due to their convective interiors, which generate strong magnetic fields. This activity can significantly impact RV measurements and, in some cases, even result in false planet detections \citep[e.g.,][]{2001A&A...379..279Q, 2014Robertson}. 

To mitigate these effects, we utilized several additional parameters extracted from the cross-correlation function (CCF) of each spectrum, alongside the RVs. These parameters, which are also available in HARPS-RVBank, include the CCF Full-Width at Half-Maximum (FWHM), Contrast, Bisector Inverse Slope (BIS), differential Line Width (dLW) and ChRomatic indeX (CRX) \citep{2018A&A...609A..12Z,2020A&A...636A..74T}. \element{H}$\alpha$ and \element{NaD} chromospheric line emission \citep{2003A&A...403.1077K} are also provided in HARPS-RVBank.

This set of quantities, usually referred to as "activity indicators," all quantify various aspects of changes in the spectrum shape that can be associated with stellar activity \citep[e.g.,][]{2001A&A...379..279Q, 2011Gomes_da_Silva, 2011A&A...535A..55D}. We used them in our analysis to further establish the orbital nature of the detected signal, as described in the following section. The photometric data available for the target stars, including observations from Hipparcos \citep{1997A&A...323L..49P}, TESS \citep{2015JATIS...1a4003R}, ASAS-SN \citep{2017PASP..129j4502K}, and ZTF \citep{2019PASP..131a8002B}, were also inspected for transit events, but yielded no noteworthy results.

\section{Analysis}
\label{sec:analysis}

{\renewcommand{\arraystretch}{1.5}%
\begin{table}[ht]
    \centering
    \caption{Stellar parameters of \object{GJ\,2126}}
    \begin{tabular}{lcc}
    \hline
    \hline
        Parameter & \object{GJ\,2126} & Ref.\\
        \hline
        RA (deg) & 257.360671 &1 \\
        DEC (deg) & -35.391731 &1\\
        $\pi$ (mas) & $26.0913 \pm 0.0174$ & 1 \\
        $\mu_\alpha$ (mas year$^{-1}$) & $-46.6689\pm 0.1227$  & 1\\
        $\mu_\delta$ (mas year$^{-1}$) & $-90.76979\pm 0.0722$ & 1\\
        \hline
        $T_{\mathrm{eff}}$ (K) & $4159\pm 129$ &  2\\
        log $g$ & $4.5270^{+0.0996}_{-0.1340}$ &  2\\
        $R (R_\odot )$ & $0.728^{+0.064}_{-0.078}$ &  2\\
        $M$ $(M_\odot)$ & $0.650^{+0.092}_{-0.075}$ &  2\\
        $L$ $(L_\odot)$ & $0.1428\pm 0.0104$ &  2\\
        ${\mathrm{[Fe/H]}}$ (dex) & $0.60\pm 0.09$ & 3 \\
        distance (pc) & $38.1290\pm 0.1171$ & 1\\
        \hline
        Sp.T. & M0V & 3\\
        B & $11.959\pm 0.016$ &  2\\
        V & $10.70\pm 0.03$ &  2 \\
        B-V & $1.259\pm 0.034 $ &  2\\
        G & $10.1574\pm 0.0028$ & 1\\
        J & $8.283\pm 0.021$ &  2 \\
        H & $7.643\pm 0.034$ &  2\\
        K & $7.557\pm 0.029$ &  2\\
    \hline
    \hline
    \end{tabular}
    \begin{minipage}{7cm} \textbf{References:} 1.~\citet{2016A&A...595A...1G, 2023A&A...674A...1G}, 2.~\citet{2019AJ....158..138S}, 3.~\citet{2019ApJ...878..134K}    
    \end{minipage}
    \label{tab:table 1}
\end{table}}

\subsection{Periodograms}

To initially identify the periodic signal in the measurements, we applied two different periodograms to the RV data: the Generalized Lomb-Scargle periodogram \citep[GLS;][]{GLSpaper} and the Phase Distance Correlation periodogram \citep[PDC;][]{zucker18, pdc_errors}.

While the GLS periodogram was selected due to its widespread use, driven by its versatility and computational efficiency, it exhibits an inherent bias toward sinusoidal signals \citep{Pinamonti17, zucker18}. This bias arises from the fact that the GLS fits a sinusoidal model to the RVs at each trial frequency. The better the model fits the data at a given frequency, the higher the likelihood that the data include a periodic signal at that frequency.

To address this limitation and improve the detection of non-sinusoidal signals, such as those associated with eccentric orbits, we employed the PDC periodogram \citep{zucker18}. The PDC periodogram quantifies the statistical dependence between the RV measurements and their phases (according to the trial periods) using the distance correlation dependence measure \citep{Szeetal2007}. Unlike the GLS periodogram, the PDC periodogram is model-independent, enabling more robust detection of a broader range of signal types, including those arising from eccentric orbits and sawtooth-like pulsation mechanisms.

\begin{figure*}
\begin{minipage}{.49\textwidth}
    {\includegraphics[width=\textwidth]{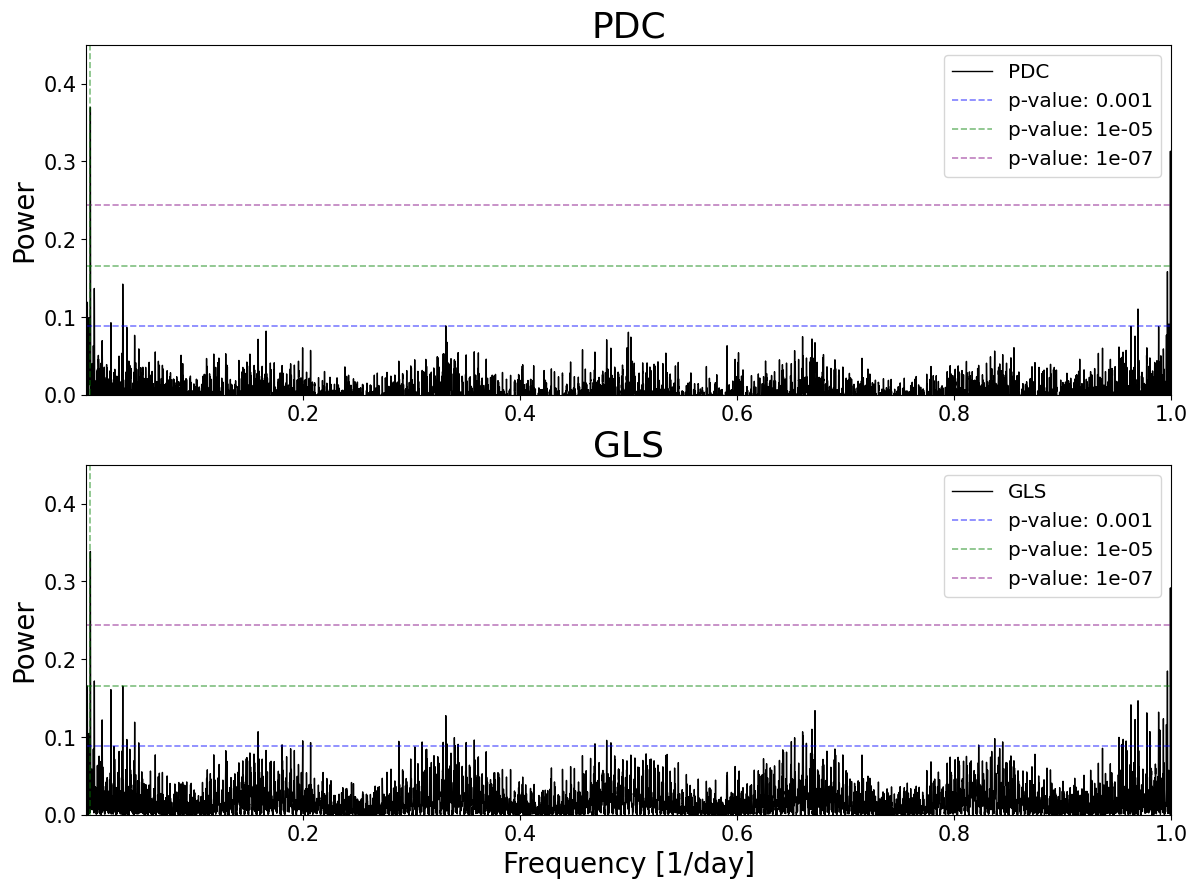}}
\end{minipage}
\hfill    
\begin{minipage}{.49\textwidth}
    {\includegraphics[width=\textwidth]{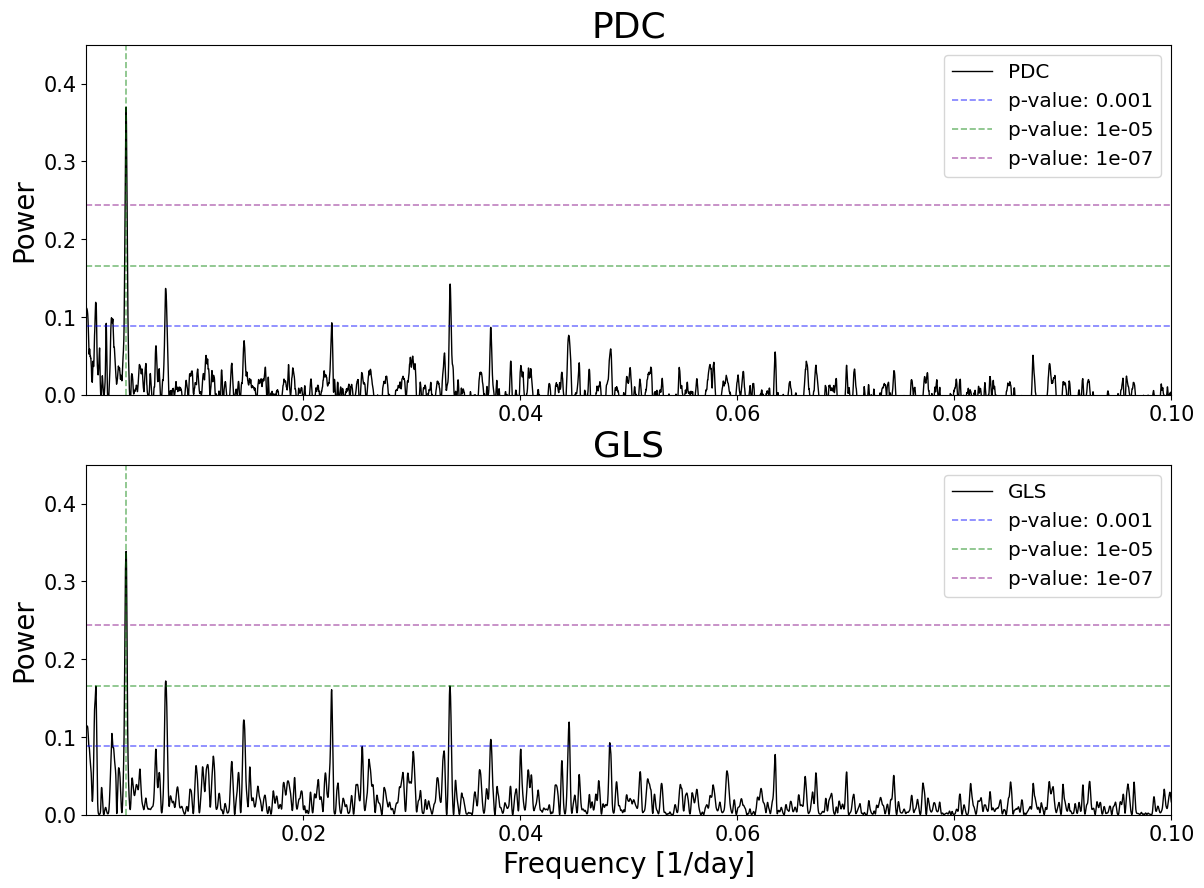}}
\end{minipage}
    \caption{PDC (top left) and GLS (bottom left) periodograms for the RV data. Also shown are the periodograms with a narrower frequency range -- PDC (top right) and GLS (bottom right). The horizontal dotted lines indicate the corresponding p-values as noted in the legend. The vertical green dotted line in each periodogram marks the detected period of $273.22$ days.}
    
    \label{fig:periodograms}
\end{figure*}

Figure~\ref{fig:periodograms} presents the periodograms for the RVs of \object{GJ\,2126}. In both periodograms, a prominent peak is visible at the frequency corresponding to a period of $\sim 273$ days.

Two additional significant peaks, though less prominent, are visible in the periodogram: one at a period of $\sim137$ days, which corresponds to a clear harmonic at half the period of the primary peak. The second, at $\sim 30$ days, is likely an artifact of the lunar cycle. This is supported by the fact that during the years 2004–2012, when most of the RV data were collected, the full moon passed within $15^{\circ}$ of the target star for many months.

\begin{figure*}
    \centering
    \includegraphics[width=2\columnwidth]{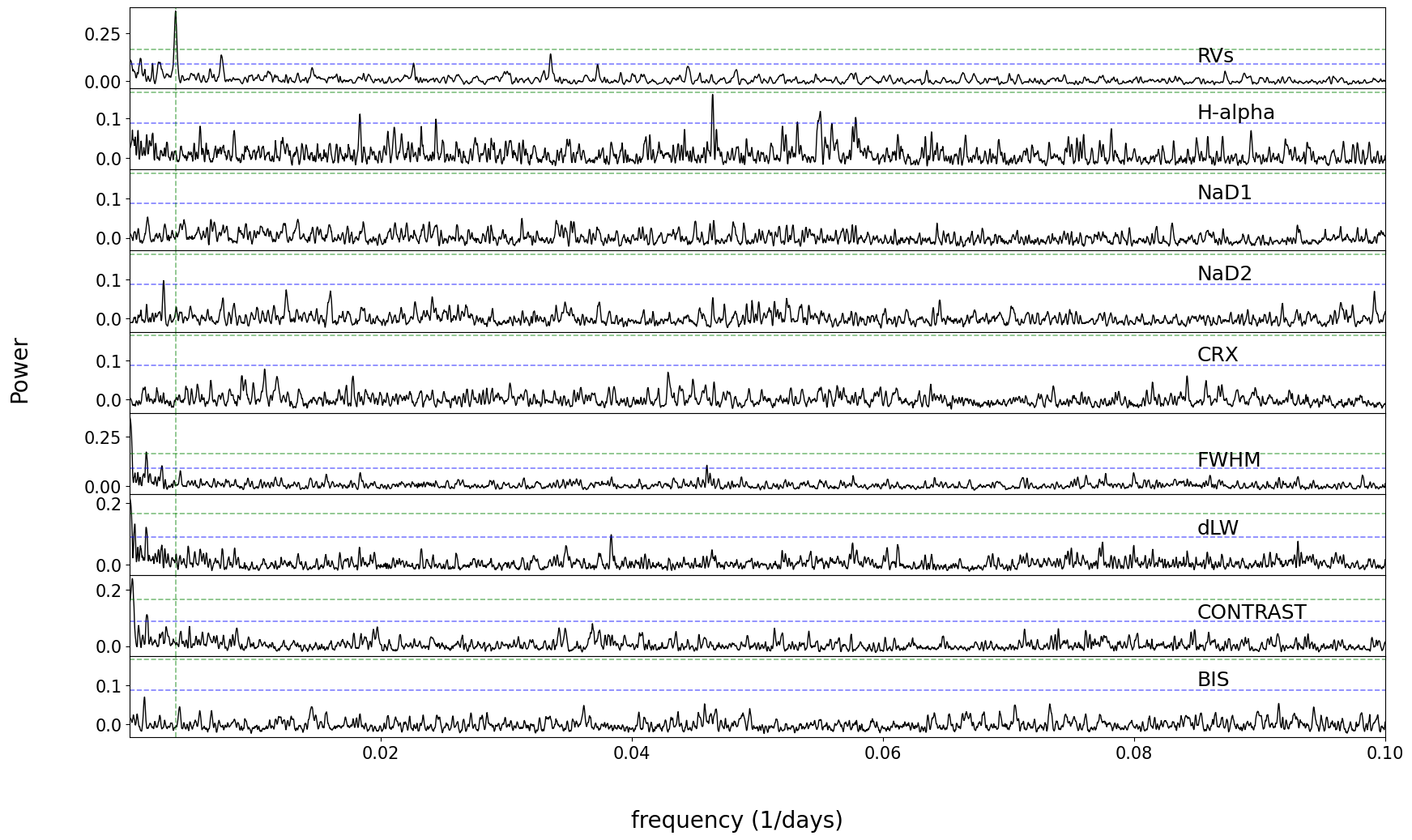}
    \caption{PDC periodograms for the RVs (top) and activity indicators (below). A green vertical line marks the detected planetary period. Blue and green horizontal lines mark the corresponding $10^{-3}$ and $10^{-5}$  p-values, respectively.}\label{fig:activity indicators}
\end{figure*}

The periodograms for the various activity indicators provided by HARPS-RVBank (see Sect.~\ref{sec:Measurements}) are presented in Fig.~\ref{fig:activity indicators}. None of the activity indicators showed a peak corresponding to the periods detected in the RV periodogram, making it unlikely that the RV periodicity is related to stellar activity or rotation. A small peak at a period of $\sim 21.5$ days seem to be present, mainly in \element{H}$\alpha$ and to a lesser extent in the FWHM, potentially linked to stellar activity or rotation at this frequency \citep{2023Mignon}.

\subsection{Orbital fitting}
\label{sec:Orbital Fitting}

We fitted a Keplerian orbit to the RVs using \texttt{Juliet} \citep{Espinoza_2019}, which employs the \texttt{radvel} code \citep{radvel} for estimating the parameters of a Keplerian model. The \texttt{emcee} sampler \citep{emcee} available in \texttt{Juliet} was used to perform a Markov Chain Monte Carlo (MCMC) sampling of the posteriors. In addition to the Keplerian elements ($P$, $T_0$, $\sqrt{e} \sin{\omega}$, $\sqrt{e} \cos{\omega}$, $K$), we fitted a separate offset term for the pre- and post-fiber upgrade datasets ($\mu_{\text{pre}}$ and $\mu_{\text{post}}$). We also included an additional white-noise jitter term for each HARPS dataset, denoted by $\sigma_{\text{pre}}$ and $\sigma_{\text{post}}$, to account for any extra noise not captured by the nominal RV error estimates.

To account for potential long-term trends in the data that could indicate the presence of outer companions, we also tested the RVs for a linear slope. The resulting posterior distribution for the slope parameter is consistent with having no slope over the observational baseline, leading us to adopt a model without a linear trend. The lack of a linear trend can provide constraints on the effects of a potential long-period companion, limiting its influence to below the measurement errors over the observation time span. Table~\ref{tab:priors} presents the priors we used for the MCMC sampling, utilizing a wide range of parameters to ensure comprehensive exploration of the parameter space.

\begin{figure*}
    \centering
    \includegraphics[width=1.9\columnwidth]{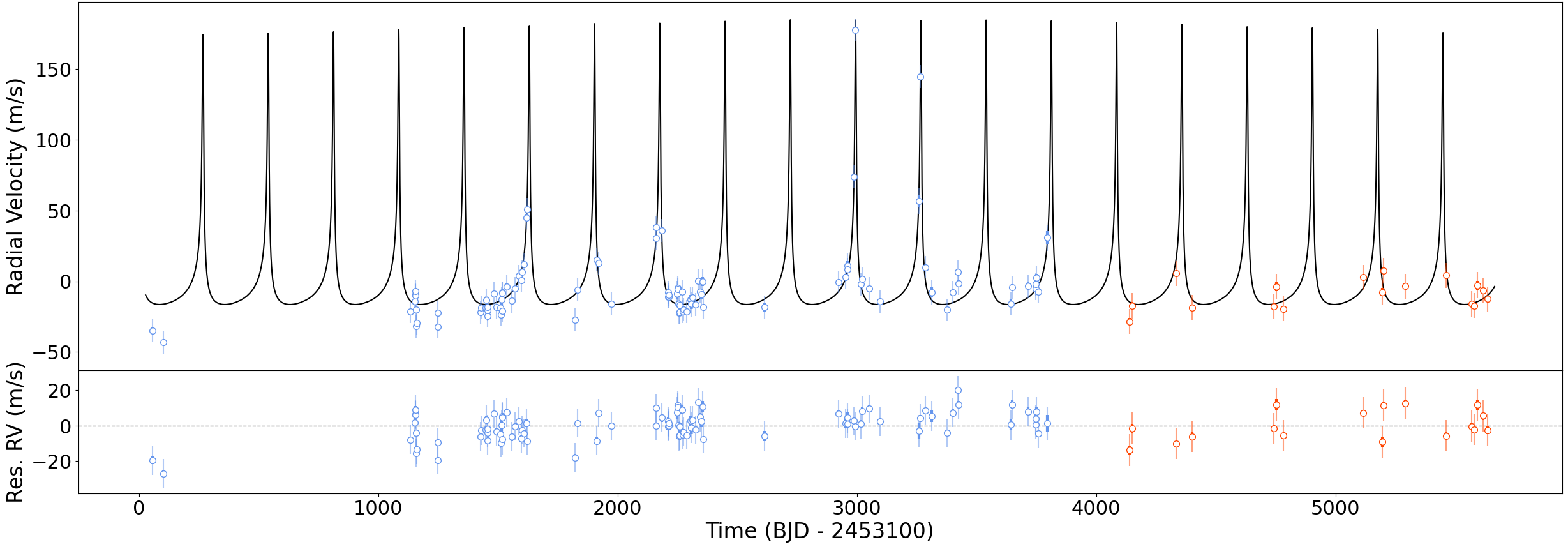}
    \caption{Best-fit Keplerian model for the full RV data (top), and corresponding residuals (bottom). Blue dots represent pre-fiber upgrade measurements, and orange dots represent post-upgrade ones.}\label{fig:full_fit}
\end{figure*}

\begin{figure*}
    \centering
\includegraphics[width=1.9\columnwidth]{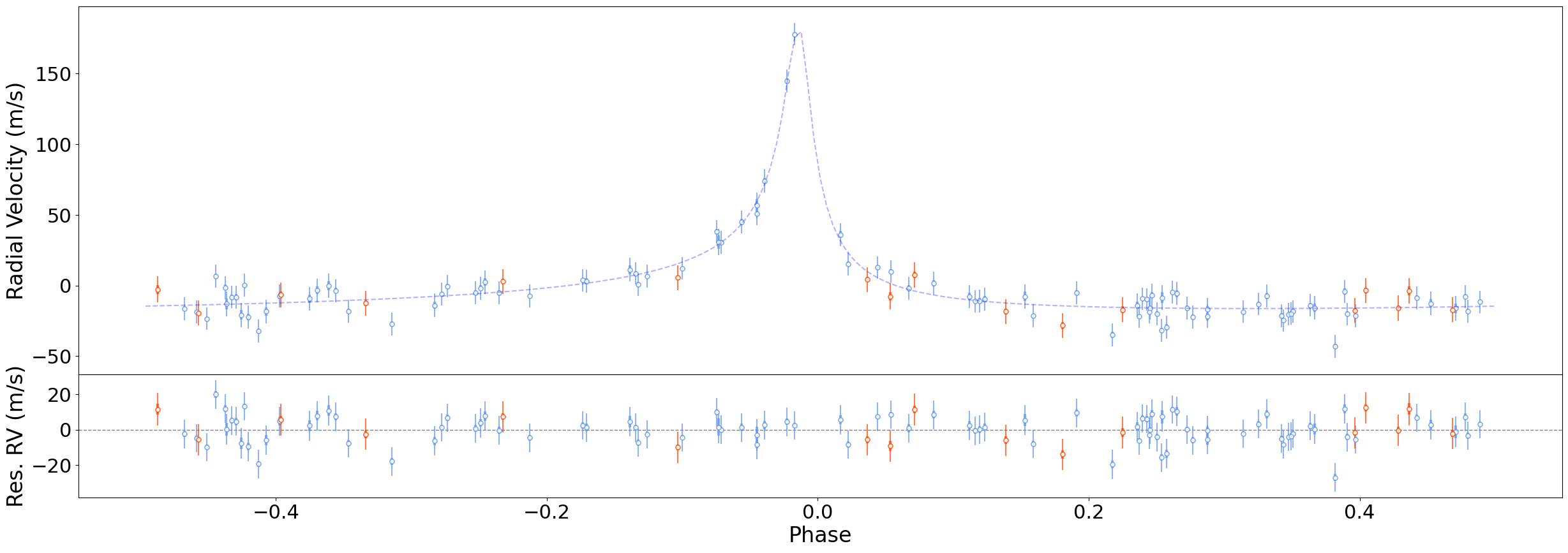}
    \caption{The RVs phase-folded according to the best-fit Keplerian model (top), and the corresponding residuals (bottom). Blue dots represent pre-fiber upgrade measurements, and orange dots represent post-upgrade ones.}\label{fig:phase folded}
\end{figure*}

\begin{figure*}
    \par\medskip
    \centering
    \includegraphics[width=0.9\linewidth]{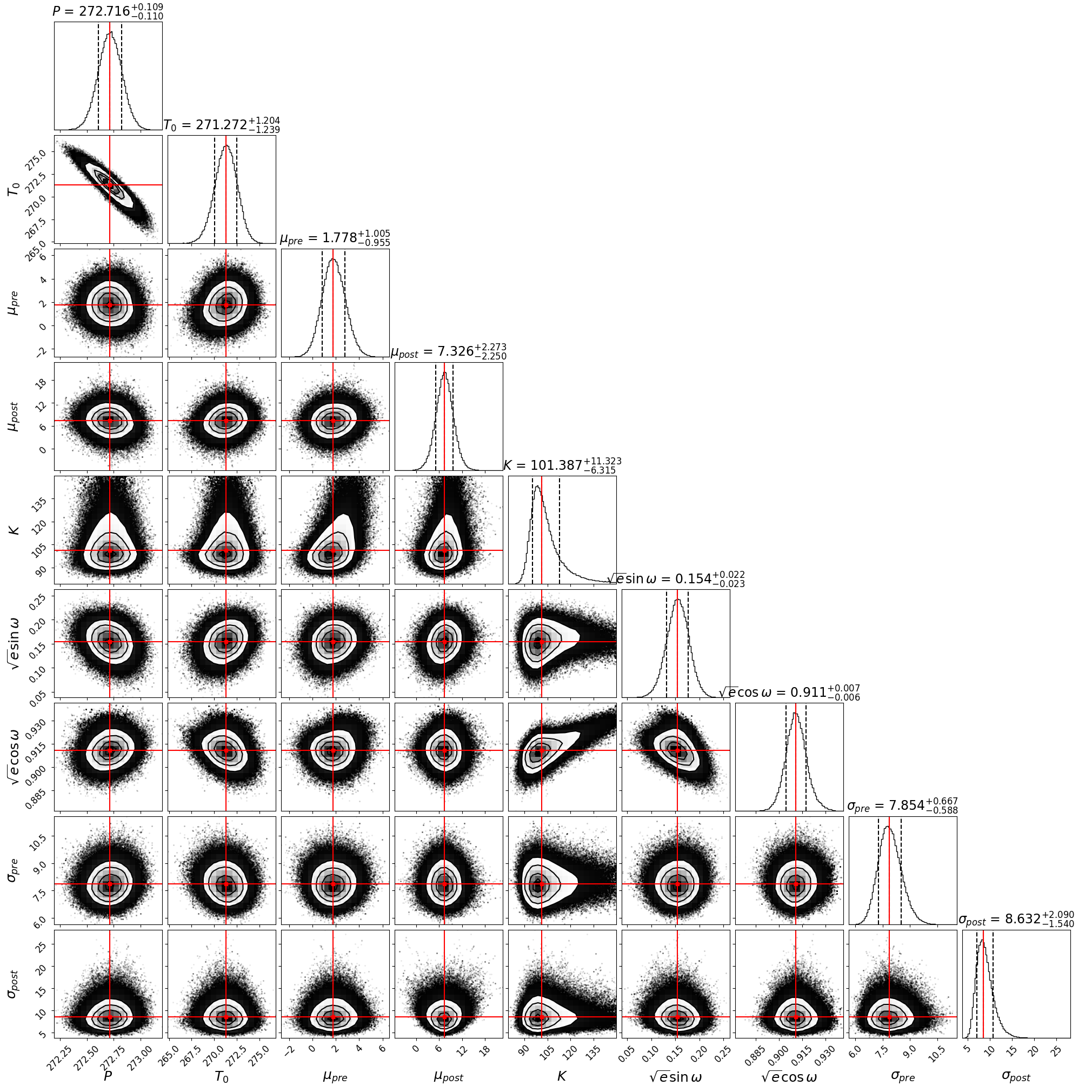}
    \caption{Corner plot for the best-fit parameters.}
    \label{fig:corner}
\end{figure*}

Based on the results of the MCMC run, we derived a best-fit Keplerian model, which is detailed in table \ref{tab:keplerian model} and illustrated in Fig.~\ref{fig:full_fit}. The keplerian elements $e$ and $\omega $ presented in table \ref{tab:keplerian model} were obtained from the fitting parameters  $\sqrt{e} \sin{\omega} = 0.15 \pm 0.02$, $\sqrt{e} \cos{\omega} =0.91 \pm 0.01$. Figure \ref{fig:phase folded} displays the phase-folded data along with the best-fit model, while Fig.~\ref{fig:corner} presents the complete corner plot from the MCMC analysis.

We also assessed the periodicity of the model residuals by subtracting the Keplerian solution from the RVs and applying both the PDC and GLS periodograms. No residual signals were detected in this analysis.

\citet{2010Anglada} and \citet{2019MNRAS.484.4230W} extensively examined the scenario in which a resonant 1:2 two-planet system with circular orbits could be misidentified as a single eccentric planet. To exclude this possibility, we experimented with multi-planetary models during the fitting process, which ultimately ruled out such configurations by providing no viable fit. Furthermore, the sufficient phase coverage, particularly around periastron, combined with the high S/N, further strengthens our confidence in the detection of this eccentric exoplanet. It is also worth noting that, according to current formation models and observational constraints, it is still uncertain whether low-mass stars are likely to host two or more giant planets \citep[see][for references]{2023MNRAS.521.3663B}.


\begin{table}[ht]
   \caption{Prior distributions used for the MCMC runs}
    \centering
    \begin{tabular}{lcc}
    \hline
    \hline
        Parameter & Units &  Distribution \\
        \hline
        \multicolumn{3}{c}{Keplerian elements}\\
        \hline
        $P$ & days & $\mathcal{N }$ $(273.0$ , $0.2)$\\
        $T_0-2453100$ & days & $\mathcal{U}$ $(100$ , $400)$\\
        $K$ & m s$^{-1}$ & $\mathcal{U}$ $(50$ , $150)$\\
        $\sqrt{e} \sin{\omega}$ & - & $\mathcal{U}$ $(-1$ , $1)$\\
        $\sqrt{e} \cos{\omega} $ & - & $\mathcal{U}$ $(-1$ , $1)$ \\
        \hline
        \multicolumn{3}{c}{Instrument-related parameters}\\
        \hline
        $\mu_{\text{pre}}$ & m s$^{-1}$ & $\mathcal{U}$ $(-20$ , $ 20)$\\
        $\mu_{\text{post}}$ & m s$^{-1}$ & $\mathcal{U}$ $(-25$ , $ 25)$\\
        $\sigma_{\text{pre}}$ & m s$^{-1}$ & $\mathcal{U}$ $(0.001$ , $ 40)$\\
        $\sigma_{\text{post}}$ & m s$^{-1}$ & $\mathcal{U}$ $(0.001$ , $ 40)$\\

    \hline
    \hline
    \end{tabular}
 
    \label{tab:priors}
\end{table}

{\renewcommand{\arraystretch}{1.5}%
\begin{table}[ht]
    \centering
    \caption{Best-fit Keplerian elements for \object{GJ\,2126\,b}}
    \begin{tabular}{lcc}
    \hline
    \hline
        Parameter & Value & Units\\
        \hline
        \multicolumn{3}{c}{Keplerian elements}\\
        \hline
        $P$ & $272.7 \pm 0.1$ & days \\
        $T_0$ & $2453371 \pm 1$ & BJD\\
        $K$ & $101^{+11}_{-6}$ & m s$^{-1}$ \\
        $e$ & $0.85 \pm 0.01$ & -\\
        $\omega $ & $10 \pm 1$ & degrees\\
        \hline
        \multicolumn{3}{c}{Instrument-related parameters}\\
        \hline
        $\mu_{pre}$ & $1.8 \pm 1.0$ & m s$^{-1}$ \\
        $\mu_{post}$ & $ 7 \pm 2$& m s$^{-1}$ \\
        $\sigma_{pre}$ & $7.9^{+0.7}_{-0.6} $ & m s$^{-1}$\\
        $\sigma_{post}$ & $8.6^{+2.1}_{-1.5} $ & m s$^{-1}$\\
         \hline
        \multicolumn{3}{c}{Derived Quantities}\\
        \hline
        $M \sin i$ & $1.3^{+0.2}_{-0.1}$ & $M_{\mathrm{Jup}}$ \\
        $a$ & $0.71 \pm 0.03$ & AU \\
    \hline
    \hline
    \end{tabular}
    \label{tab:keplerian model}
\end{table}}

\section{Discussion}
\label{sec:discussion}

In this paper we report the detection and orbital characterization of \object{GJ\,2126\,b}, a highly eccentric ($e = 0.85$) Jupiter-mass exoplanet orbiting its low-mass host star with a period of $272.7$ days. Its unique properties place it in a relatively sparse region of the detected exoplanet population, offering valuable insights for current planetary formation and evolution models.

As discussed in Sect.~\ref{sec:Measurements}, \object{GJ\,2126}, is widely referred to in the literature as an M-dwarf star. Only a relatively small number of eccentric exoplanets have been detected around M-dwarf stars, despite their importance for various open questions. These include dynamically unexplained cases of high eccentricities in systems where circularization was expected \citep[e.g.,][]{2014ApJ...796...32S}, and correlations between eccentricities and various stellar parameters \citep[see][]{2023Sagear}. Another key question is the habitability of M-dwarf star systems, where tidal heating caused by high eccentricity is expected to eliminate the possibility of life-supporting conditions \citep{Barnes_2013}. This discovery contributes a valuable piece of data to this exoplanet demographics and it can be used to refine observing strategies and models \citep[e.g.,][]{2021Sabotta}.

With its inclination remaining unknown, it could be argued that the detected companion is significantly more massive than the suggested $1.3 M_{\mathrm{J}}$ and in fact not planetary in its nature. For the companion to be classified within the brown dwarf regime, its orbital orientation would need to be nearly face-on ($i \lesssim 6 ^\circ$). This is highly unlikely given the distribution of observed inclinations \citep[e.g.,][and references therein]{2022Bertaux}.

Furthermore, we ruled out this possibility based on astrometric measurements from both Hipparcos and Gaia, which exclude the existence of such a massive companion \citep{2019Kervella, 2022Kervella}. Additionally, the Renormalized Unit Weight Error (RUWE) value for the star from Gaia third Data Release (DR3) is approximately one, strongly indicating the absence of an astrometric signal of a massive companion \citep{gaia_mult, 2023A&A...674A..32B}. Given the characteristics of \object{GJ\,2126\,b}, the detection of such a companion should have been well within Gaia capabilities \citep{2014ApJ...797...14P, 2014Sozzetti}.

The reported exoplanet calls for an explanation that can account for its high eccentricity. Since planets are assumed to form within protoplanetary disks, their initial orbits should have been circular or close to circular, as the disk gaseous viscosity tends to dampen any orbital eccentricity \citep[e.g.,][]{2013Bitsch}. Gravitational interactions between planets and the disk, where eccentricity increases due to significant gaps in the disk \citep{2001A&A...366..263P}, are only relevant for the eccentricity of very massive planets exceeding five Jupiter masses \citep{2006A&A...447..369K}.

The Lidov–Kozai effect \citep{1979Mazeh, 2016Naoz} is a potential mechanism that could account for the observed eccentricity by inducing oscillations in the orbit inclination and eccentricity through gravitational interactions with an additional, long-period massive companion in the system. However, as discussed earlier, astrometric measurements from Hipparcos and Gaia rule out the existence of such a companion, excluding this possibility. The absence of a long-term trend in the data, discussed in Sect.~\ref{sec:Orbital Fitting}, further supports this conclusion.

A scattering event induced by a stellar flyby also seems unlikely due to the relatively close-in orbit of the planet. As demonstrated by \citet{2004AJ....128..869Z}, a stellar flyby with an impact parameter of several hundred Astronomical Units (AU) could significantly affect planets with longer-period orbits ($a \geq 50 \mathrm{AU}$). However, planets with shorter orbits would not experience substantial perturbations from such encounters \citep{2019Bancelin}. While closer flybys might occur during early evolution stages within a star cluster, the resulting eccentricity of the planets would likely be damped over time by the surrounding protoplanetary disk \citep{2014Picogna}.

The most likely explanation we propose for the observed high eccentricity is scattering through planet-planet interactions. Recent simulations indicate that such interactions can lead to extreme eccentricity values \citep{2019A&A...629L...7C}. Specifically, this process involves multiple young planets with moderate eccentricities forming in suitable locations and interacting with one another. As some planets are ejected from the system, angular momentum transfer can increase the remaining planets eccentricities \citep{2018AJ....156..192W}. \citet{2008ApJ...686..621F} demonstrated that these scattering events are most effective when the interacting planets have comparable masses, which provides further insight into the potential evolutionary history of \object{GJ\,2126}. It is also plausible that the described process is part of high-eccentricity tidal migration, an evolutionary pathway that may explain the origins of hot Jupiters \citep[see][]{2018ARA&A..56..175D}.

\begin{acknowledgements}
We are grateful to the anonymous referee for their insightful comments, which helped improve the manuscript. This research was supported by the ISRAEL SCIENCE FOUNDATION (grant No.\ 1404/22) and the Israel Ministry of Science and Technology (grant No.\ 3-18143).
This work has made use of data from the European Space Agency (ESA) mission {\it Gaia} (\url{https://www.cosmos.esa.int/gaia}), processed by the {\it Gaia} Data Processing and Analysis Consortium (DPAC, \url{https://www.cosmos.esa.int/web/gaia/dpac/consortium}). Funding for the DPAC has been provided by national institutions, in particular the institutions participating in the {\it Gaia} Multilateral Agreement.
The analyses done for this paper made use of the code packages: \texttt{NumPy} \citep{numpy}, \texttt{SciPy} \citep{2020SciPy}, \texttt{Juliet} \citep{Espinoza_2019}, \texttt{radvel} \citep{radvel} and \texttt{SPARTA} \citep{SPARTA2020}\footnote{The PDC periodogram and other periodograms used in this work, along with spectrum analysis tool, are all available as part of the SPARTA package \citep{SPARTA2020}, at \url{https://github.com/SPARTA-dev}.}.
\end{acknowledgements}

\bibliographystyle{aa}
\bibliography{main}

\end{document}